\documentstyle[12pt]{article}
\def\ltap{\raisebox{-.4ex}{\rlap{$\sim$}} \raisebox{.4ex}{$<$}}

\def\journal{\topmargin .3in    \oddsidemargin .5in
        \headheight 0pt \headsep 0pt
        \textwidth 5.625in 
\textheight 8.25in 
        \marginparwidth 1.5in
        \parindent 2em
        \parskip .5ex plus .1ex         \jot = 1.5ex}
\journal

\def\ra{\rightarrow}
\begin{document}
\begin{titlepage}
\begin{center}
March 22, 1999      \hfill    LBNL-43000\\
\vskip .5in

{\large \bf Quantum corrections from nonresonant $WW$ scattering}
\footnote
{This work was supported by the Director, Office of Energy
Research, Office of High Energy and Nuclear Physics, Division of High
Energy Physics of the U.S. Department of Energy under Contract
DE-AC03-76SF00098.}

\vskip .5in

Michael S. Chanowitz\footnote{Email: chanowitz@lbl.gov}

\vskip .2in

{\em Theoretical Physics Group\\
     Ernest Orlando Lawrence Berkeley National Laboratory\\
     University of California\\
     Berkeley, California 94720}
\end{center}

\vskip .25in

\begin{abstract}

An estimate is presented of the leading radiative corrections to low 
energy electroweak precision measurements from strong nonresonant $WW$ 
scattering at the TeV energy scale.  The estimate is based on a novel 
representation of nonresonant scattering in terms of the exchange of 
an effective scalar propagator with simple poles in the complex energy 
plane.  The resulting corrections have the form of the corrections 
from the standard model Higgs boson with the mass set to the unitarity 
scale for strong $WW$ scattering.

\end{abstract}

\vskip .2in
{\it To be published in the Lev Okun festschrift volume of 
Physics Reports C}
\end{titlepage}

\renewcommand{\thepage}{\roman{page}}
\setcounter{page}{2}
\mbox{ }

\vskip 1in

\begin{center}
{\bf Disclaimer}
\end{center}

\vskip .2in

\begin{scriptsize}
\begin{quotation}
This document was prepared as an account of work sponsored by the United
States Government. While this document is believed to contain correct
 information, neither the United States Government nor any agency
thereof, nor The Regents of the University of California, nor any of their
employees, makes any warranty, express or implied, or assumes any legal
liability or responsibility for the accuracy, completeness, or usefulness
of any information, apparatus, product, or process disclosed, or represents
that its use would not infringe privately owned rights.  Reference herein
to any specific commercial products process, or service by its trade name,
trademark, manufacturer, or otherwise, does not necessarily constitute or
imply its endorsement, recommendation, or favoring by the United States
Government or any agency thereof, or The Regents of the University of
California.  The views and opinions of authors expressed herein do not
necessarily state or reflect those of the United States Government or any
agency thereof, or The Regents of the University of California.
\end{quotation}
\end{scriptsize}

\vskip 2in

\begin{center}
\begin{small}
{\it Lawrence Berkeley National Laboratory is an equal opportunity employer.}
\end{small}
\end{center}

\newpage

\renewcommand{\thepage}{\arabic{page}}
\setcounter{page}{1}
\noindent {\it \underline {Prologue}}

I first met Lev Okun at the 1976 ``Rochester'' conference held in the 
USSR, in Tbilisi, Georgia.  Sakharov was under strong attack by the 
government for his human rights activities and was originally not 
invited but was permitted to attend after he protested the lack of an 
invitation to the Soviet Academy.  Understandably even those Soviet 
physicists who were sympathetic to Sakharov and his ideas were 
cautious about associating with him during the meeting.  While there 
may well have been others I did not observe, to me it was remarkable 
to see one Soviet physicist who did not hesitate to stroll openly with 
Sakharov on the streets of Tiblisi.  This was of course Okun.  His 
behavior then demonstrated the same simple idealism and courage that 
is reflected now in the decision he has taken since the dissolution of 
the USSR to remain in Moscow, to preserve the unique physics 
environment at ITEP, when he could easily have accepted more 
comfortable positions outside of Russia.

Not unrelated to his moral character is the clarity, depth and 
humanity with which Okun practices physics.  This gives me a selfish 
reason for submitting the work presented here: I would like to have 
his view of it.  It has a plausible conclusion reached by a 
strange method and raises questions I do not understand.  It is based 
on a representation of an exactly unitary model of nonresonant $WW$ 
scattering in terms of an effective scalar propagator with simple 
poles in the complex energy plane.  The method was applied and 
verified for tree approximation amplitudes and is used here to 
estimate quantum corrections.

\noindent {\it \underline {Introduction} }

The electroweak symmetry may be broken by weakly coupled Higgs bosons 
below 1 TeV or by a new sector of quanta at the TeV scale that 
interact strongly with one another and with longitudinally polarized 
$W$ and $Z$ bosons.  Precision electroweak data favors the first 
scenario\cite{okun_ew,ewwg}, but the conclusion is not definitive, 
because the relevant quantum corrections are open to contributions 
from many forms of new physics.  Occam's (an archaic spelling of 
Okun's?)  razor favors the simplest interpretation, which assumes that 
the only new physics contributing significantly are the quanta that 
directly form the symmetry breaking condensate.  In that case the data 
do favor weak symmetry breaking by Higgs scalars.  But nature may have 
dealt us a more complicated hand, with other, probably related, new 
physics also contributing to the radiative corrections.  Then the 
precision data tells us nothing about the symmetry breaking sector --- 
unless we can ``unscramble'' the different contributions, which in 
general we do not know how to do --- and implementation of the Higgs 
mechanism by strong, dynamical symmetry breaking remains a 
possibility.  The nature of the symmetry breaking sector can only be 
established definitively by its direct discovery and detailed study in 
experiments at high energy colliders.

Strong $WW$ scattering is a generic feature of strong, dynamical 
electroweak symmetry breaking.\cite{mcmkg2} The longitudinal 
polarization modes $W_{L}$ scatter strongly above 1 TeV because the 
enforcement of unitarity is deferred to the mass scale of the heavy 
quanta that form the symmetry breaking condensate.  To the extent that 
QCD might be a guide to dynamical symmetry breaking we expect the 
$a_{00}$ partial wave to smoothly saturate unitarity between 1 and 2 
TeV. Like the SM (standard model) Higgs boson, nonresonant strong $WW$ 
scattering would also contribute to the low energy radiative 
corrections probed in precision electroweak measurements.  This note 
presents an estimate of those corrections, based on a novel 
representation of nonresonant strong $WW$ scattering as an 
effective-Higgs boson exchange amplitude.

Strong $WW$ scattering models are customarily formulated in R-gauges.  
The effective-Higgs representation allows them to be reexpressed gauge 
invariantly and, in particular, in unitary gauge.\cite{noewa1,noewa2} 
It applies to the leading $s$-wave amplitudes with $I=0,2$.  The 
effective-Higgs representation has a significant practical advantage: 
it predicts the experimentally important transverse momentum 
distributions of the final state quark jets and the $WW$ diboson in 
the collider process $qq \rightarrow qqWW$, which cannot be obtained 
from the conventional method based on the effective $W$ 
approximation.\cite{ewa} The method has been verified numerically for 
tree amplitudes\cite{noewa1} and gauge (i.e., BRST) invariance has 
been demonstrated.\cite{noewa2}
   
The K-matrix model is a useful model of strong $WW$ scattering which 
smoothly extrapolates the $WW$ low energy theorems\cite{mcmkg2,let} in 
a way that exactly satisfies elastic unitarity.  The effective-Higgs 
representation of the K-matrix model has a surprisingly simple 
form: the singularities of the propagator are simple poles in the 
complex $s$ plane, like an elementary scalar.  It is then easy to 
compute the contribution to the $W$ and $Z$ vacuum polarization 
tensors from which the ``oblique'' corrections\cite{pt,ab} are 
obtained.

The final result for the oblique parameters  
$S$ and $T$ is like the SM Higgs contribution 
with $m_{H}$ replaced by a combination of the unitarity scales 
for strong  scattering in the $I=0,2$ channels, 
determined in turn by the low energy theorems 
as noted in \cite{mcmkg2}.  $S$ and $T$ are given by
$$
S={1\over 18\pi}
\left[{\rm ln}\left(
{16\pi v^{2}\over \mu^{2}}\right) +
               {1\over 2}\ {\rm ln}\left({32\pi v^{2}\over \mu^{2}}\right) 
 \right]
\eqno(1)
$$
$$
T= {-1\over 8\pi\ {\rm cos}^{2}\theta_{W}}
       \left[{\rm ln}\left({16\pi v^{2}\over \mu^{2}}\right)  +
                {1\over 2}\ {\rm ln}\left({32\pi v^{2}\over \mu^{2}}
                \right) \right]
\eqno(2)
$$
where $v^{2}=(\sqrt{2}G_{F})^{-1}$, $\theta_{W}$ is the weak 
interaction mixing angle and $\mu$ is the reference scale. For $\mu = 
1$ TeV the corrections are $S \simeq 0.036$ and $T\simeq -0.11$.
Similar results follow from the cut-off nonlinear sigma model when the 
unitarity scales are used for the cutoffs.\cite{nls}

In the following sections I review the K-matrix model, derive the 
effective scalar propagator, deduce the oblique corrections, raise  
some theoretical issues, and finally discuss the physical 
interpretation of the result. 
 
\noindent {\it \underline {K-matrix model for $WW \rightarrow ZZ$}}

In the SM the Higgs sector contribution to $WW \rightarrow ZZ$ is 
given by just the $s$-channel Higgs pole. Therefore we use the 
K-matrix model for $WW \rightarrow ZZ$ to abstract the effective-Higgs 
propagator. The model is summarized in this section.

As is conventional we use the ET\cite{et} (equivalence theorem) to 
define the model in terms of the unphysical Goldstone bosons, 
$w^{\pm}$ and $z$. 
Partial wave unitarity is conveniently formulated as 
$$
	{\rm Im}\ {1 \over a_{IJ}} = -1.  \eqno(3)
$$	
The K-matrix model is constructed to satisfy the low energy theorems 
and partial wave unitarity. It is defined by
$$
{1 \over a^{K}_{IJ}} = {1\over R_{IJ}} - i    \eqno(4)
$$
where $R_{IJ}$ are the real threshold amplitudes that 
follow from the low energy theorems,
$$
R_{00} = {s \over 16\pi v^{2}}   \eqno(5a)
$$
$$
R_{20} = {-s \over 32\pi v^{2}}. \eqno(5b)
$$
The corresponding $s$-wave T-matrix amplitudes are
$$
{\cal M}_{I}^{K}(s)= 16\pi a^{K}_{I0} \eqno(6)
$$
for $I=0,2$. Finally the $ww\rightarrow zz$ amplitude is 
$$
{\cal M}^{K}(w^{+}w^{-}\ra zz) = 
            {2\over 3}({\cal M}_{0}^{K} - {\cal M}^{K}_{2}) 
\eqno(7)
$$

\noindent {\it \underline {Effective-Higgs propagator}}

To obtain the effective-Higgs propagator we ``transcribe'' the 
K-matrix model from R-gauge to U-gauge.\cite{noewa1,noewa2} The heart 
of the matter is to find the contribution of the symmetry-breaking 
sector in U-gauge, which encodes the dynamics specified in the 
original R-gauge formulation of the model.  This is accomplished using 
the ET as follows.

Suppose that the longitudinal gauge boson modes scatter strongly. 
At leading order in the weak gauge coupling $g$ we write 
the amplitude $W_{L}^{+}W_{L}^{-}\ra ZZ$ as a sum 
of gauge-sector and Higgs-sector terms,
$$
{\cal M}_{\rm Total} = {\cal M}_{\rm Gauge} + {\cal M}_{\rm SB}
\eqno(8)
$$
where SB denotes the symmetry breaking (i.e., Higgs) sector.
Gauge invariance ensures that the contributions to ${\cal M}_{\rm 
Gauge}$ that grow like $E^{4}$ cancel, leaving a sum that grows like 
$E^{2}$, given by 
$$
{\cal M}_{\rm Gauge}= g^2{E^2 \over \rho m_W^2} + 
         {\rm O}(E^{0},\ g^{4})
\eqno(9)
$$
where $\rho = m_W^2/({\rm cos}^2\theta_W m_Z^2)$.  The neglected terms 
of order $E^{0}$ and of higher order in $g^{2}$ include the 
electroweak corrections to the leading strong amplitude.

The order $E^2$ term in equation (9) is the residual 
``bad high energy behavior'' that is cancelled by the 
Higgs mechanism. It is also precisely the 
low energy theorem amplitude,
$$
{\cal M}_{\rm LET} = {s \over \rho v^2} = {\cal M}_{\rm Gauge} 
+ {\rm O}(s^{0},\ g^{4})
\eqno(10)
$$
using $m_W=gv/2$ and $s=4E^2$. Eqs. (8) and  (9) may 
be used to derive the low energy theorem  without invoking the 
ET.\footnote{
If the symmetry breaking force is strong, the quanta of the symmetry 
breaking sector are heavy, $m_{SB} \gg m_W$, and decouple in gauge 
boson scattering at low energy, ${\cal M}_{SB} \ll {\cal M}_{\rm 
Gauge}$.  Then the quadratic term in ${\cal M}_{\rm Gauge}$ dominates 
${\cal M}_{\rm Total}$ for $m_W^2 \ll E^2 \ll m_{SB}^2$, which 
establishes the low energy theorem without using the 
ET.\cite{let}}

Now consider an arbitrary strong scattering model, designated  
as model ``X'', formulated in the usual way in an R-gauge in terms 
of the unphysical Goldstone bosons, ${\cal M}^{\rm X}_{\rm 
Goldstone}(ww\rightarrow zz)$.  The total gauge boson amplitude is 
gauge invariant and the ET tells us that for $E\gg m_{W}$ it is 
approximately equal to the Goldstone boson amplitude, i.e.,
$$
{\cal M}^{\rm X}_{\rm Total}(W_{L}W_{L}) 
\simeq {\cal M}^{\rm X}_{\rm Goldstone}(ww)
\eqno(11)
$$
in the same approximation as eq. (9).
Eq. (8) holds in any gauge. Specifying U-gauge 
we combine it with eqs. (9-11) to obtain the U-gauge Higgs sector 
contribution for model X, 
$$
{\cal M}_{\rm SB}^{\rm X}(W_LW_L) = 
{\cal M}^{\rm X}_{\rm Goldstone}(ww) - {\cal M}_{\rm LET}.
\eqno(12)
$$

The preceding result applies to any strong scattering amplitude.  
Now we specialize to  $s$-wave $WW\ra ZZ$ scattering and 
use eq.(12) to obtain 
an effective-Higgs propagator with standard ``Higgs''-gauge boson 
couplings.  Neglecting $m_{W}^{2}\ll s$ and higher orders in $g^{2}$ 
as always, the effective scalar propagator is 
$$
P_X(s) = - {v^2 \over s^2}{\cal M}_{\rm SB}^{\rm X}(W_LW_L) 
\eqno(13)
$$
Eqs.(10) and (12) with $\rho = 1$ then imply
$$
P_X(s) = -{v^2 \over s^2}{\cal M}^X_R(ww) + {1\over s}
              \eqno(14)
$$
The term 1/$s$, corresponding to a massless scalar, comes from ${\cal 
M}_{\rm LET}$ in eq.  (12).  It ensures good high energy behavior  
while the other term in eq. (14) 
expresses the model dependent strong dynamics.

Finally we substitute the 
K-matrix amplitude, eq. (7), into eq. (14) to obtain the effective 
propagator for the K-matrix model as the sum of two simple poles
$$
P_{K}= {2\over 3}\left(
          {1\over s -m_{0}^{2}} + 
          {1\over 2}{1\over s -m_{2}^{2}}\right)    \eqno(15)
$$
where $m_0$ and $m_{2}$ are 
$$
m_{0}^{2}= -16\pi i v^{2}    \eqno(16)
$$
and
$$
m_{2}^{2}= +32\pi i v^{2}.    \eqno(17)
$$
It is surprsing to find such a simple expression involving only 
simple poles. It is not surprising that the poles are far from the 
real axis since they describe nonresonant scattering. 
Interpreted heuristically as Breit-Wigner poles they correspond to 
resonances with widths twice as big as their masses. 

\noindent {\it \underline {Oblique corrections}}

The oblique corrections are evaluated from the vacuum polarization 
diagrams that in the SM include the Higgs boson.\cite{pt}  In place of 
the SM propagator, $P_{\rm SM}=1/(s-m_{H}^{2})$, we substitute $P_{K}$ 
from eq.  (15).  Where the SM contribution depends on the log of the 
Higgs boson mass, $L_{SM}={\rm ln}(m_{H}^{2}/\mu ^{2})$, we now find 
instead the combination $L_{K}$,
$$
  L_{SM}={\rm ln}\left({m_{H}^{2}\over \mu ^{2}}\right)\ \ \ra \ \ 
  L_{K}={2\over 3}\ {\rm ln}\left({m_{0}^{2} \over \mu ^{2}}\right) + 
  {1\over 3}\ {\rm ln}\left({m_{2}^{2}\over \mu ^{2}}\right) 
     \eqno(18)
$$
where $m_{0,2}$ are complex masses defined in eqs. (16-17).

The results quoted in eqs.  (1-2) follow from the usual expressions 
for $S,T$ where we use the real part of $L_{K}$ in place of $L_{SM}$,
$$
S= {{\rm Re}\left(L_{K}\right) \over 12\pi} 
   \eqno(19)
$$
and 
$$
T= {-3\ {\rm Re}\left(L_{K}\right) 
            \over 16\pi\ {\rm cos}^{2}\theta_{W}}
   \eqno(20)
$$
The imaginary part of $L_{K}$ is an artifact which we discard; it 
results from the fact that our approximation neglects the $W$ mass, as 
in any application of the ET. At $q^{2}=0$, where the oblique 
corrections are computed, there is no contributution to the 
imaginary part of the vacuum polarization from the relevant diagrams. 

Combining the $I=0$ and $I=2$ terms in eq. (18) we have 
$$
{\rm Re}\left(L_{K}\right) =
      {\rm ln}\left({2^{1/3}16\pi v^{2} \over \mu ^{2}}\right). 
\eqno(21)
$$
Evaluating eq. (21) we find that the oblique correction from the 
K-matrix model is like that of a Higgs boson with mass 2.0 TeV.

\noindent {\it \underline {Questions}}

The $I=2$ component of the effective propagator has peculiar 
properties, perhaps due to the fact that for the 
$I=2$ channel we are representing $t$- and $u$-channel dynamics by an 
effective $s$-channel exchange.  The minus sign in the $I=2$ low 
energy theorem, eq.  (5b), which may be thought of as arising from the 
identity $t+u=-s$, leads to interesting differences between the $I=0$ 
and $I=2$ components of the effective propagator $P_{K}$.

First, the $I=2$ component of the effective scalar propagator has a 
negative pole residue, which would correspond to a unitarity violating 
ghost if it described an asymptotic state (which it does not).  In 
fact the sign is required to {\em ensure} unitarity, since it is 
needed to cancel the bad high energy behavior of the gauge sector 
amplitude which has a negative sign in the $I=2$ channel.  In eq.  
(15) for $P_{K}$ the $I=2$ pole appears with a positive sign because 
of a second minus sign from the isospin decomposition, eq.  (7).  
Neither pole of the effective propagator has a negative (ghostly) 
residue.  In any case the amplitude is exactly unitary by 
construction.

The sign difference between the pole positions, $m_{0}^{2}$ and 
$m_{2}^{2}$ in eqs.  (16) and (17), may also be traced to the 
phases of the low energy theorems in eq.  (5).  The position 
of $m_{0}^{2}$ on the negative imaginary axis of the complex $s$ plane 
corresponds to poles in the fourth and second quadrants of the complex 
energy plane, consistent with causal propagation as in the 
conventional $m^{2}-i\epsilon$ prescription.  But the position of 
$m_{2}^{2}$ on the positive imaginary axis corresponds to poles in the 
first and third quadrants of the complex energy plane.  This would 
imply acausal propagation if the poles are on the first sheet but not 
if they are on the second sheet.  Working in the limit of massless 
external particles as we are it is not apparent on which sheet they 
occur.\footnote{I thank Henry Stapp for a discussion of this point.}

I conclude that the sign of the pole residue arising from the 
$I=2$ amplitude is not problematic but that the implications of the 
pole position requires better understanding. 
\newpage

\noindent {\it \underline {Physical interpretation}}

We have used a convenient representation of the K-matrix model to 
estimate the low energy radiative corrections from strong $WW$ 
scattering.  The result that the corrections are like those of a 
Higgs boson with mass at the unitarity scale is plausible and agrees 
with an earlier estimate using the cut-off nonlinear sigma 
model.\cite{nls} The estimate establishes a `default' radiative 
correction from the strongly coupled longitudinal gauge bosons in 
theories of dynamical symmetry breaking.  In general there will be 
additional contributions from other quanta in the symmetry 
breaking sector.  Those contributions are model dependent as to 
magnitude and sign.  In computing their effect it is important 
to avoid double-counting contributions that are dual to the 
contribution considered here.

Current SM fits to the electroweak data prefer a light Higgs boson 
mass of order 100 GeV with a 95\% CL upper limit that I will 
conservatively characterize as $\ltap 300$ GeV.\cite{ewwg} Since the 
corrections computed here are equivalent to those of a Higgs boson 
with a mass of 2 TeV, they are excluded at 4.5 standard deviations.  
Therefore there must be additional, cancelling contributions to the 
radiative corrections from other quanta in the theory if strong $WW$ 
scattering occurs in nature.  This would not require fine-tuning 
although it would require a measure of serendipity.

There are good reasons for the widespread view that a light 
Higgs boson is likely and for the popular designation of SUSY 
(supersymmetry) as The People's Choice.  But SUSY also begins to 
require a measure of serendipity\cite{price} to meet the increasing 
lower limits on sparticle and light Higgs boson masses.  While the 
community of theorists has all but elected SUSY, the question is not 
one that can be decided by democratic processes.  At the end of the 
day only experiments at high energy colliders can tell us what the 
symmetry breaking sector contains.  Collider experiments, 
particularily those at the LHC, should be prepared for the full range 
of possibilities, including the capability to measure $WW$ scattering 
in the TeV region.
\vskip .2in
\noindent {\it Acknowledgements:} I wish to thank David 
Jackson, and Henry Stapp for helpful discussions.  This work was 
supported by the Director, Office of Energy Research, Office of High 
Energy and Nuclear Physics, Division of High Energy Physics of the 
U.S. Department of Energy under Contracts DE-AC03-76SF00098.

\end{document}